\documentclass{raa}            

\usepackage{graphicx,times}             
\usepackage{threeparttable}
\begin{document}

   \title{Analysis of the 3C445 Soft X-ray Spectrum as Observed by Chandra high-energy gratings}

   \volnopage{Vol.0 (20xx) No.0, 000--000}      
   \setcounter{page}{1}          

   \author{F. T. Dong
      \inst{1}
   \and S. H. Shao
      \inst{1}
   \and Y. Cheng
      \inst{1}
   \and J. L. Zeng
      \inst{2}
   }

   \institute{Technological Vocational College of Dezhou,
              Dezhou 251200, Shandong, China; {\it dongfutong@sohu.com} \\
        \and
             College of Science, National University of Defense Technology,
               Changsha 410073, Hunan, China \\
   }

   \date{}

\abstract{We present a detailed analysis of the soft X-ray emission of 3C445 using
 an archival Chandra HETG spectrum. Highly-ionized H- and He-like Mg, Si and S lines,
 as well as a resolved low-ionized Si K$\alpha$ line, are detected in the high resolution spectrum.
 The He-like triplets of Mg and Si are resolved into individual lines, and the calculated R
 ratios indicate a high density for the emitter. The low values of the G ratios indicate
 the lines originate from collisionally ionized plasmas. However, the detection of a resolved narrow
Ne X RRC feature in the spectrum seems to prefer to a photoionized environment. The spectrum is subsequently
 modelled with a photoionization model, and the results are compared with that of a collisional
 model. Through a detailed analysis on the spectrum, we exclude a collisional origin for these
 emission lines. A one-component photoionization model provides a great fit to the emission features.
 The best-fit parameters are log$\xi$ = $3.3^{+0.4}_{-0.3}$ erg cm s$^{-1}$,
 $n_{H}$ = $5^{+15}_{-4.5}\times10^{10}$ cm$^{-3}$ and $N_{H}$ = $2.5^{+3.8}_{-1.7}\times10^{20}$ cm$^{-2}$.
 According to the calculated high density for the emitter, the measured velocity widths of the
 emission lines and the inferred the radial distance
 (6 $\times$ $10^{14}$ - 8 $\times$ $10^{15}$ cm), we suggest the emission lines originating from
 matter locate in the broad line region (BLR).
}

   \authorrunning{F. T. Dong, S. H. Shao, Y. Cheng \& J. L. Zeng}            
   \titlerunning{Analysis of the 3C445 Soft X-ray Spectrum}  

   \maketitle

%
%
\section{Introduction}           
\label{sect:intro}

With the development of high resolution observations of the active galactic nucleis (AGNs),
more and more emission features are observed in the soft X-ray spectrum.
Emission features revealed by the high resolution spectrum, e.g. the RRC (radiative
recombination continua) and the He-like triplets, are normally identified as emission from
highly ionized H- and He-like ions of astrophysical abundant elements (e.g. NGC 3783,
Kaspi et al.~\cite{k01}; NGC 1068, Kinkhabwala et al.~\cite{k02}; NGC 4151, Armentrout et al.~\cite{a07}).
Modeling these features with theoretical models has been proved to be a powerful tool
to diagnose the physical process of generating these emission features (i.e. collisional
ionization or photoionization), and to infer the physical properties of the line
emitting gas, e.g. electron density, column density, and their distribution around the central
region (Young et al.~\cite{y01}, Bianchi et al.~\cite{b06}, Kraemer et al.~\cite{k08}).

3C445 (z=0.0562) is classified as a Broad Line Radio Galaxy (BLRG) with a
FRII morphology (Kronberget al.~\cite{k86}). For the radio-loud objects, the central engines
have been poorly studied so far due to the relative rarity and far distance of
these sources. The soft X-ray emission lines, however, provide a useful way
of interpreting the inner environment of these sources. To date several BLRG
objects have been observed with soft X-ray emission lines, for instance:
3C33  (Torresi et al.~\cite{t09}), 3C234 (Piconcelli et al.~\cite{p08}) and
3C445 (Sambruna et al.~\cite{s07}, Reeves et al.~\cite{r10}, Braito et al.~\cite{b11}).

3C445 is an ideal target for investigating the nature of the central engines
of the radio-loud objects. High resolution observations have revealed a
multitude of soft X-ray emission features from H- and He-like ions in this
object. The XMM-Newton RGS spectra provided the first detection of the O VII
and O VIII emission lines (Grandi et al.~\cite{g07}). These lines are considered most likely
from a warm photoionized gas. However, the relative low resolution and short
exposure time restrict an accurate measurement of the line parameters.
The subsequent Chandra LETG spectrum of 3C445
detected and resolved emission lines from H- and He-like O, Ne, Mg, and Si ions (Reeves et al.~\cite{r10}).
The spectrum gives the first detailed measurement of the soft X-ray emission
of this object. Based on the emission features from O VII and O VIII,
Reeves et al.~(\cite{r10}) inferred that the emission comes from photoionized gas
which probably locates in the broad line region (BLR).

The most recent Chandra HETG spectrum provides some distinct emission features
in the soft X-ray band from previous observations, e.g. resolved Mg XI and
Si XIII triplets. These new characteristics may provide more accurate constraints
on the physical states of the line emitting gas. As far as we know, no research
on the soft X-ray emission has been published in the literature based on these
data. Consequently, in this paper, we make a detailed research on the soft X-ray
emission of 3C445 based on the spectrum observed by Chandra HETG. Our
goal is to analyze the soft X-ray emission features in the spectrum and to make a detailed
comparison with that obtained with the Chandra LETG spectrum (Reeves et al.~\cite{r10}),
in order to better understand the physical properties of the soft X-ray emitting gas.

\section{Data reduction}
\label{sect:data}

The High Energy Transmission Grating (HETG) onboard Chandra observed 3C445 from
25 July 2011 to 9 August 2011. These data was obtained from the Chandra Public
Data Archive\footnote{http://cxc.harvard.edu/cda/}, with four sequences of
observations: 13305, 13306, 13307 and 14194. The data were reduced with the
standard CIAO 4.9 tools (Fruscione et al.~\cite{f06}). In order to apply the most updated
calibrations, we ran the Chandra reprocessing script to create new spectra files.
Only the first order dispersed spectra were considered for both the MEG and HEG, and
the $\pm$1 orders for each grating were subsequently combined for each sequence.
As no significant variability was found between the four sequences, we combined the
spectra from all the four sequences to yield a single 1st order spectrum for each of
the MEG and HEG, with a total exposure time of 413 ks. The spectra was used in the
energy range of 1.2 - 5 keV for MEG and 3 - 8 keV for HEG. Below 1.2 keV, the signal
to noise of the MEG data is very low and no significant emission line feature
was detected. While above 8 keV, the HEG spectrum is not actually detected
above the background. The mean count rates were 0.0117 counts/s and 0.0127
counts/s for MEG (1.2 - 5 keV) and HEG (3 - 8 keV) respectively.

The resultant time-averaged spectra were subsequently analyzed using the spectra
fitting tool $Sherpa$ (Freeman et al.~\cite{f01}). The spectra were binned at 2 different levels,
i.e. $\Delta$ $\lambda$ = 0.005 {\AA} and 0.01 {\AA} for HEG and MEG respectively,
or at the FWHM of their resolution. The former finer binning was used for the spectral
fitting and line identification, while the latter was used only for plotting purpose.
We used the $C$-statistic (Cash~\cite{c79}) for the subsequent spectral fit,
as the counts per bin drops below N $<$ 10 in some bins in the soft X-ray band.
The error bars quoted throughout the paper correspond to the $90\%$ confidence
interval (i.e. $\Delta{C}$ = 2.706 for one interesting parameter).

In this paper, we also used the Chandra LETG data for comparison purpose.
The observation was performed from 2009 September 25 to October 3, with Chandra
ACIS-S for a total exposure time of 200 ks. These data has been analyzed by Reeves
et al.(~\cite{r10}). We reduced the data following the standard procedure as discribed
above, and obtained a time-averaged spectrum with mean count rates of $\sim$ 0.0132 counts/s
in the 0.5 - 9 keV band.

\section{Line Identification}
\label{sect:lines}
Initially, we concentrated on the analysis of the MEG spectrum to search for the
soft X-ray emission lines and measure their properties. As an overall preview
of the emission line features, Figure 1 shows the MEG spectrum fluxed against a broken power-law.
Since there are no significant emission lines are resolved above 2.4 keV,
the spectrum was only plotted in the energy range of $\sim$ 1.2 - 2.4 keV.
The spectrum shown in Figure 1 clearly exhibits several strong
emission line structure in the AGN rest frame, as confirmed by previous studies
with the high resolution Chandra LETG and XMM-Newton spectrum
(Reeves et al.~\cite{r10}, Braito et al.~\cite{b11}). The expected emission features in the
observed frame are marked in the figure. Most of these lines are expected
from H- and He-like Mg, Si and S ions. There is also a strong emission feature at
$\sim$ 1.65 keV which may be identified as the Si K$_{\alpha}$ line.

  \begin{figure}
   \centering
   \includegraphics[width=10cm, angle=0]{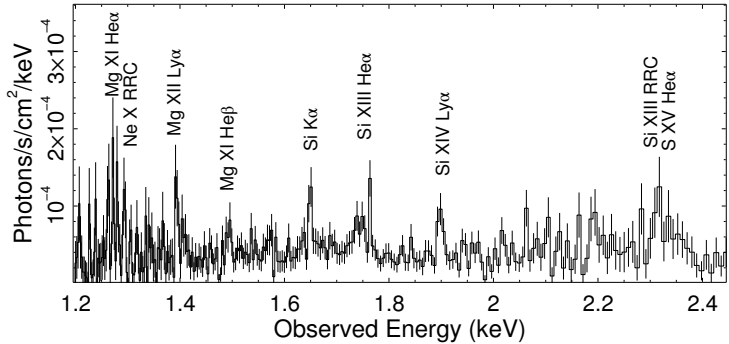}
   \caption{The fluxed Chandra MEG spectra of 3C445 in the emission line energy band.}
   \label{Fig1}
   \end{figure}

Due to the low quality of data below 1.2 keV, no emission feature can be detected in the present
spectrum from low-Z ions, e.g. the O VII, O VIII, Ne IX lines as observed in the LETG spectrum
(Reeves et al.~\cite{r10}). However, the high resolution and long exposure time guarantee the
MEG spectrum to resolve some emission features that have never been detected in the LETG
observation. Figure 2 shows a direct comparison of the LETG spectrum with the MEG in the
Mg and Si line band. Apparently, the MEG spectrum shows a better resolution
of the emission lines, specifically for the He-like triplets of Mg and Si which
appear to be resolved into their individual lines in the MEG spectrum, while these lines
are barely resolved in the LETG spectrum. In addition, the MEG spectrum also resolved a narrow
feature around 1.29 keV in the observed frame, which may be identified as the RRC from Ne X.
On the contrary, no such feature is detected in the LETG spectrum.

  \begin{figure}
   \centering
   \includegraphics[width=8cm, angle=0]{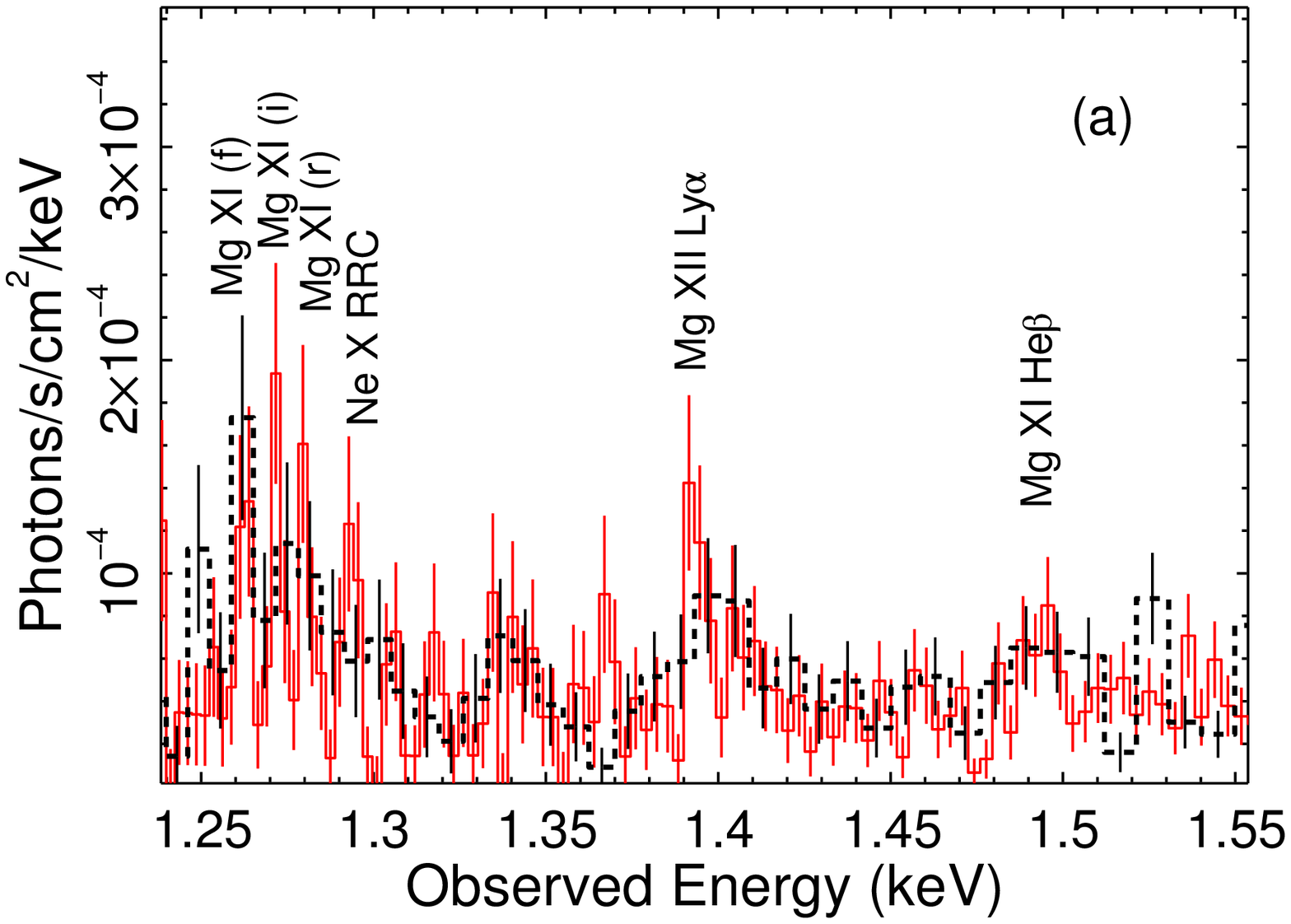}
   \includegraphics[width=8cm, angle=0]{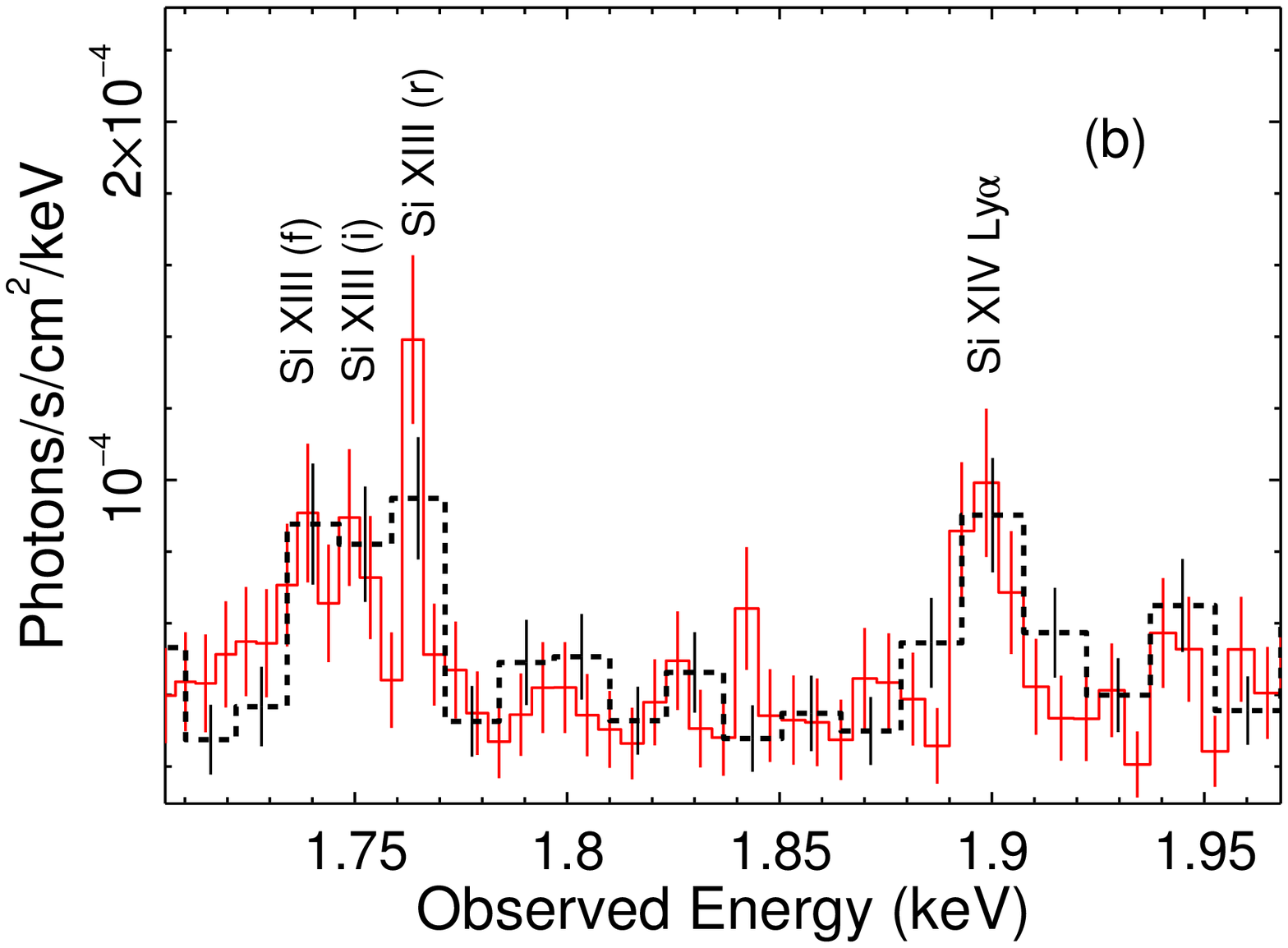}
   \caption{Chandra MEG (red solid line in the electronic version) and LETG
   (black dashed line in the electronic version) spectra folded against a broken power-law continuum.
   Panel (a) shows the spectra in the Mg band. Panel (b) shows the spectrum in the Si
   band.}
   \label{Fig2}
   \end{figure}

In order to extract parameters for these emission features as accurately as possible,
we fitted the spectrum using a 'local fits' method as described in Guainazzi \& Bianchi(~\cite{gb07})
and Nucita et al.(~\cite{n10}).
In principle, this method provides more accurate measurement of the line parameters than the 'global
fits' method if the continuum is very complex, by removing any uncertainties as to how the continuum
is modeled. Previous X-ray studies of 3C445 revealed the continuum is heavily absorbed by complex
absorbing components (Sambruna et al.~\cite{s07}, Braito et al.~\cite{b11}).
Besides a strong cold reflection component is also required to predict the hard excess above 10 keV
and the strong Fe K$\alpha$ line at $\sim$ 6.4 keV. Thus the complexity of the continuum may affect the
measurement of the line parameters using the 'global fits' method.

Upon modeling each emission feature, the spectrum is divided into intervals of 0.1 keV wide
which well covers the selected emission feature.
The continuum can be well modeled with a simple power-law in such a narrow energy region.
The lines were fitted using a Gaussian profile with the centroid energy, line width and line flux left
free to vary, while the RRC profile was modeled with a recombination emission edge.
Note upon modeling the He-like triplets of Mg XI and Si XIII, we assumed that the widths of the
three lines of each ion were identical and tied these values in the resulting model. As for
the He-like triplet of S XV, no individual line appears to be resolved in the spectra. Thus
this component was fitted with a single Gaussian profile. The Galactic absorption was accounted
for in the spectral analysis by using the neutral absorption model $wabs$ with the column
density $N_H$ = 4.6 $\times$ $10^{20}$ cm$^{-2}$ for 3C445 (Dickey \& Lockman~\cite{dl90}).
The measured line fluxes, widths, equivalent widths, centroid energies and their statistical
significance are listed in Table 1.

\begin{table}
\begin{center}
\begin{threeparttable}
\caption[]{Summary of Line Parameters detected on the MEG Spectrum.}\label{Tab:1}
\begin{tabular}{ccccccc}
  \hline\noalign{\smallskip}
Line ID & Line Energy (eV)\tnote{a} & Flux ($10^{-7}$ photons/cm$^{2}$/s)\tnote{b} & $\sigma$ or kTe (eV)\tnote{c}
& EW (eV)\tnote{d} & $\Delta{C}$\tnote{e} &  E$_{exp}$ (eV)\tnote{f} \\
 \hline\noalign{\smallskip}
Mg XI He$\alpha$ (r) & $1352^{+1}_{-1}$ & $4.8^{+3.3}_{-2.5}$ & $<$ 2  & $13.9^{+9.5}_{-7.2}$    & 12 &1352\\
Mg XI He$\alpha$ (i) & $1343^{+1}_{-1}$ & $5.8^{+3.5}_{-2.6}$ &$--$& $11.6^{+6.9}_{-5.3}$ & 19 &1344\\
Mg XI He$\alpha$ (f) & $1334^{+1}_{-1}$ & $5.1^{+3.6}_{-2.8}$ &$--$& $11.5^{+8.2}_{-6.4}$ & 11 &1331\\
Ne X RRC             & $1365^{+1}_{-2}$ & $4.1^{+3.1}_{-2.3}$ & $<$ 4  & $9.3^{+7.1}_{-5.2}$  & 11 &  1362\\
Mg XII Ly$\alpha$    & $1472^{+2}_{-1}$ & $6.7^{+3.8}_{-3.1}$ &$2.0^{+1.7}_{-0.8}$ & $17.3^{+9.2}_{-7.5}$  & 18 &1472\\
Mg XI  He$\beta$     & $1578^{+3}_{-3}$ & $4.2^{+3.0}_{-2.5}$ &$3.6^{+3.4}_{-3.6}$ &  $10.5^{+7.6}_{-5.3}$   &10 &1579\\
Si K$\alpha$         & $1742^{+2}_{-1}$ & $7.2^{+3.3}_{-2.7}$ &$1.9^{+1.8}_{-1.3}$ &  $15.8^{+7.3}_{-6.0}$  & 25 &1744\tnote{g}\\
Si XIII He$\alpha$ (r)  & $1863^{+1}_{-1}$ & $6.7^{+2.8}_{-2.3}$ & $<$ 3   &  $15.8^{+6.7}_{-5.4}$  & 35    &1865\\
Si XIII He$\alpha$ (i)  & $1847^{+4}_{-2}$ & $3.1^{+2.5}_{-1.8}$ &$--$ & $7.2^{+6.0}_{-4.3}$   & 10  &1854\\
Si XIII He$\alpha$ (f)  & $1837^{+2}_{-3}$ & $4.0^{+2.9}_{-2.1}$ &$--$ &  $9.4^{+6.8}_{-4.8}$  &14 &1839\\
Si XIV Ly$\alpha$   & $2005^{+3}_{-2}$ & $7.6^{+4.3}_{-3.0}$ & $2.5^{+3.3}_{-2.0}$ &  $19.6^{+11}_{-7.7}$   & 24 &2006\\
S XV He$\alpha$    & $2451^{+12}_{-14}$ & $19.4^{+14.6}_{-13.3}$ & $14.0^{+15.6}_{-13.9}$ & $36.9^{+27.8}_{-25.3}$  & 11 & 2447 \\

\noalign{\smallskip}\hline

\end{tabular}
\begin{tablenotes}
\item[a]{The measured line centroid energy.}
\item[b]{The measured line fluxes.}
\item[c]{The widths of the detected lines or temperature of RRC feature.}
\item[d]{Equivalent widths of the lines.}
\item[e]{Improvement in $C$-statistic after adding the feature to the model.}
\item[f]{The expected line energies taken from the CHIANTI database (Dere et al.~\cite{d01}).}
\item[g]{The value is calculated using the flexible atomic code (FAC) (Gu~\cite{g03}).}
\end{tablenotes}
\end{threeparttable}
\end{center}
\end{table}

Most of the signatures are detected at a confidence level $> 99\%$, i.e. $\Delta{C}$ $>$ 11.3
for 3 parameters or $\Delta{C}$ $>$ 9.21 for 2 parameters. Specifically, the Si XIII He-like
resonance, the Si XIV Ly$\alpha$ and the Si K$\alpha$ lines are detected with $\Delta{C}$ $>$ 21.1,
which corresponds to the confidence level $>$ 99.99$\%$. The strong Si K$\alpha$ line is evidently
from different region in the central engine from the highly ionized lines, due to its low ionization
state. Indeed this line most probably originates from reflection off a Compton-thick mirror (see Section 5).

Comparing the measured centroid energies with the expected line energies, we found that most of
the line centroids are consistent with the expected values in the quoted error range, especially
for the strongest and most isolated lines, such as Mg XII Ly$\alpha$ and Si XIV Ly$\alpha$. This
indicates the line emitter is most likely stationary relative to the local AGN. Indeed evidence
of zero velocity shift for the soft X-ray emitter has also been suggested by the Chandra LETG
spectrum (Reeves et al.~\cite{r10}).

As discussed in Braito et al.~(\cite{b11}), the flux state is rather constant for 3C445, we could
compare the present line fluxes with those measured with the Chandra LETG spectrum (Reeves et al.~\cite{r10}).
Figure 3 shows the flux differences of Mg and Si lines between the MEG and LETG spectra.
Good agreements are achieved for the H-like Mg XII Ly$\alpha$ and Si XIV Ly$\alpha$ lines,
i.e. $\sim$ 10$\%$ difference for Mg XII Ly$\alpha$ between the two observations, and $<$ 10$\%$
for Si XIV Ly$\alpha$. The He$\alpha$ (i)+(r) fluxes are consistent within $<$ 30$\%$ differences
between the two observations. However, the differences are relatively large
for the forbidden lines in the He-like triplets, e.g. the flux of the Mg XI He$\alpha$ (f) line in
LETG is more than twice that of MEG. The difference is probably because individual lines in the He-like
triplets of Mg XI and Si XIII are poorly resolved in the LETG spectrum as shown in Figure 2.

  \begin{figure}
   \centering
   \includegraphics[width=8cm, angle=0]{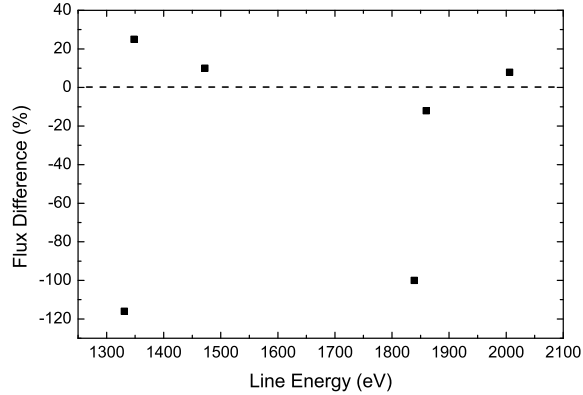}
   \caption{Flux difference of Mg and Si lines between the MEG and LETG spectra.
   Note the He$\alpha$ (i) and (r) lines of Mg XI and Si XIII
   were not resolved in the LETG spectrum, thus we compared the total flux
   (i.e. (i) + (r)) in each triplet between the two observations.}
   \label{Fig3}
   \end{figure}

The line widths of the two prominent H-like Mg XII Ly$\alpha$ and Si XIV Ly$\alpha$
lines were well determined. The Mg XII Ly$\alpha$ line was resolved with a width of
$\sigma = 2.0^{+1.7}_{-0.8}$ eV and the Si XIV Ly$\alpha$ with $\sigma = 2.5^{+3.3}_{-2}$ eV,
corresponding to FWHM widths of $1000^{+800}_{-400}$ km/s and $900^{+1200}_{-700}$ km/s
respectively. While only upper-limits were obtained for the intrinsic velocity widths for the
He-like Mg XI and Si XIII lines, i.e. $\sigma$ $<$ 2 eV and $\sigma$ $<$ 3 eV (corresponding
to FWHM widths of $<$ 1100 km/s and $<$ 1200 km/s) respectively. Comparing the soft X-ray
line FWHM with that of the H$_{\beta}$ line can potentially give a direct indication of the location
of the soft X-ray line-emitting region relative to the optical BLR.
We used the method described in Shu et al.(~\cite{s10,s11}).
Figure 4 shows the FWHM ratios of the Mg XII Ly$\alpha$ and Si XIV Ly$\alpha$ lines with that of the 
H$_{\beta}$ line (FWHM $\sim$ 3000 km/s, Osterbrock et al.~\cite{o76}, Eracleous \& Halpern~\cite{eh04}).
We found at the two-parameter 99$\%$ confidence level both of the FWHM ratios
were consistent with 1. This indicates the soft X-ray line gas may have the same
location as the BLR gas.

  \begin{figure}
   \centering
   \includegraphics[width=8cm, angle=0]{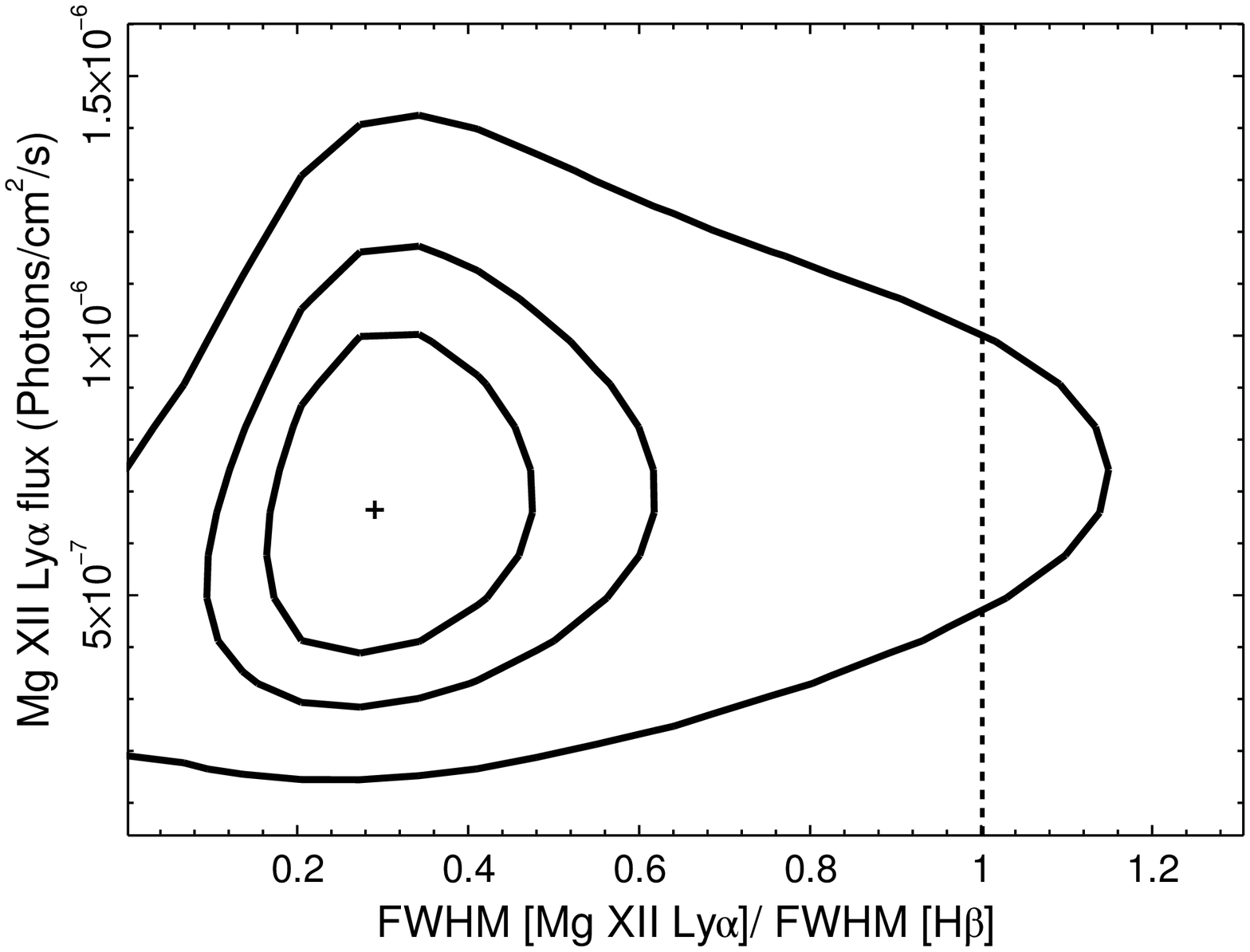}
   \includegraphics[width=8cm, angle=0]{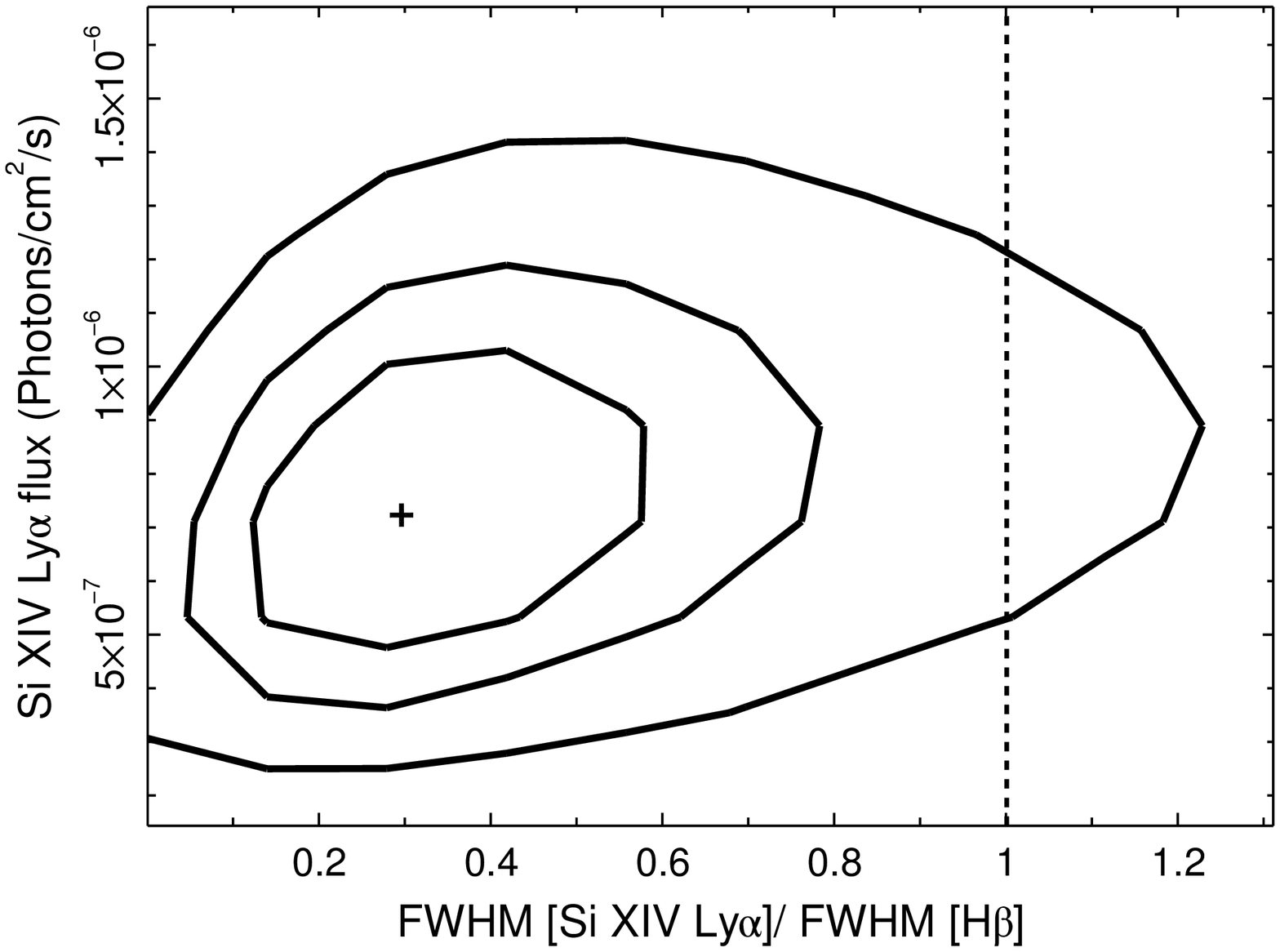}
   \caption{Joint 68$\%$, 90$\%$, and 99$\%$ confidence contours of the line flux versus
   the ratio of the line FWHM to the H$_{\beta}$ FWHM for Mg XII Ly$\alpha$ and Si XIV Ly$\alpha$ respectively. 
   The vertical dashed line corresponds to the FWHM ratio equals to 1.}
   \label{Fig4}
   \end{figure}

\subsection{Radiative Recombination Continua}

The MEG spectrum detected a distinct Ne X RRC feature that can be used for temperature
diagnostic for the plasmas. Although a feature presents near the expected energy of
Si XIII RRC (see Figure 1), it is heavily blended with the surrounding emission (S XV He$\alpha$)
for us to be able to resolve the width, and hence it is not used here.
The RRC feature is formed by a free electron recombined directly into an ion.
The energy of the free electron is sensitive to the electron temperature of the emitting gas,
therefore RRC feature will have different structures for plasmas with different temperature.
RRC features originating in hot plasmas are broad and blurred, while the features are narrow
and prominent for warm photoionized plasmas.
Thus electron temperature determined from the RRC width
can indicate either a photoionization or collisional-ionization environment of the line emitting
gas (Liedahl \& Paerels~\cite{lp96}, Liedahl~\cite{l99}).

Using the Ne X RRC, we tried to measure the electron temperature of the line emitting
gas. However the feature was found to be relatively weak with $\Delta$$C$ = 11 for 3 parameters (Table 1),
which means this feature is only detected at $>$ 98$\%$ confidence. Nonetheless, an upper limit was
obtained for the RRC width of $<$ 4 eV, implying a temperature of $kT_e$ $< 4 \times 10^4$ K for
the emitting gas. This limit is consistent with the temperature derived from the Chandra LETG
spectra using the O VII and O VIII RRCs ($kT_e \sim 3 \times 10^4$ K; Reeves et al.~\cite{r10}).
The low temperature provides an evidence of photoionization for the plasmas, as it would be
insufficient for the high-Z elements (Mg, Si ,S) to be ionized to H- and He-like ions in collisionally
ionized plasmas.

\subsection{He-like Triplets}
For the two significant He-like triplets of Mg XI and Si XIII, we presented measurements of the
standard plasma diagnostic ratios R = f/i and G = (i+f)/r (Gabriel \& Jordan~\cite{g69},
Gabriel \& Jordan~\cite{g73}, Porquet \& Dubau~\cite{pd00}), where f, i, and r are the
forbidden, intercombination, and resonance line fluxes respectively.
The line ratios provide a diagnostic of the electron density and temperature
in hot plasmas (Gabriel \& Jordan~\cite{g69}, Porquet et al.~\cite{p01}).
In photoionized condition, Porquet \& Dubau~(\cite{pd00}) found that the R ratio
is sensitive to the electron density of the plasma, and the G ratio is sensitive to the
electron temperature of the plasma and is a good indicator of the ionization mechanism of
the line emitting gas.

The values are R = $0.9^{+1.1}_{-0.5}$ and $1.3^{+2.3}_{-0.7}$, and G = $2.3^{+2.7}_{-1.1}$
and $1.1^{+0.8}_{-0.5}$ for Mg XI and Si XIII respectively. Comparing
these values with Figure 8 in  Porquet \& Dubau~(\cite{pd00}), the R ratio for
Mg XI implies an electron density in the range of $n_e$ = $10^{12}$ - $10^{14}$ cm$^{-3}$,
and an upper limit of $n_e$ $<$ $10^{14}$ cm$^{-3}$ was obtained using Si XIII. All of the
values imply a high electron density for the line emitting gas. This is consistent
with measurements from the O VII triplet in the Chandra LETG spectrum, with the electron
density $n_e$ $>$ $10^{10}$ cm$^{-3}$ (Reeves et al.~\cite{r10}). However, extracting detailed
values of density is complicated by the fact that a strong UV
radiation field can mimic the effect that a relatively high density has on the
intercombination and forbidden line strengths (Porter \& Ferland~\cite{p07}).

The low values of the G ratio from both Mg XI and Si XIII indicate a collisionally
ionized plasma. However, this is contrast to the narrow Ne X RRC feature which
prefers a photoionized environment. In fact, the G ratio is no longer suitable for
plasma diagnostic in this case. As discussed in Kinkhabwala et al.~(\cite{k02}),
in photoionized plasmas, a significant contribution of photoexcitation would raise
the intensity of the resonance transition drastically, and yields the G ratio acts
more like that in collisionally ionized case. Moreover, optical depth in the
resonance line will severely affect the values of the G ratio, yielding the G ratio
a function of the column density (Porter \& Ferland~\cite{p07}). Thus we cannot tell
the difference between collisional-ionization and photoionization only with the G
ratios. To further investigate this problem, we modeled the emission lines with a collisionally
ionized model and compared the result with that using a photoionized model in the next section.

\section{Modeling The Lines}
\label{sect:model}
In modeling the emission lines, we employed a more physical model to fit the
overall spectra, using both of the MEG and HEG data. The continuum was modeled
with a partial covering model that has been used with previous LETG and Suzaku
data (Reeves et al.~\cite{r10}, Braito et al.~\cite{b11}), in the form of
$wabs$$\times$(($pow1$+$reflion$)$\times$$zwabs$$\times$$zpcfabs$+$pow2)$.
Where $pow1$ is an absorbed power-law representing the main radiation component
from the central source, $pow2$ represents part of the main radiation scattered into our line
of sight by the ambient material, and $reflion$ is a
reflection component representing the Compton mirror detected in this object.
$zwabs$ and $zpcfabs$ is the intrinsic absorption component with one fully covering
the central radiation and the other one partially covering it. All of the components
are modified by the local galactic absorber $wabs$ with the column density of
$N_H$ = 4.6 $\times$ $10^{20}$ cm$^{-2}$. A cross normalization was adopted between the
HEG and MEG spectra, and the value was always found to be consistent with 1
($1.1^{+0.1}_{-0.1}$ for the MEG spectrum).

\subsection{Collisionally Ionized Model}

Initially, in order to test the ionization mechanism of the plasma, we fitted
the emission lines with a collisional model using the $MEKAL$ code
(Mewe et al.~\cite{m86},  Liedahl \& Osterheld~\cite{lo95}). In this model,
the abundances were fixed to solar values and the redshift was fixed to the value of
3C445 (z = 0.0562). The free parameters were the plasma temperature, the
Hydrogen density and the normalization. This model was incorporated into the
continuum model given above to fit the overall (HEG + MEG) spectra, and
produced an acceptable fit with the overall fit statistic $C$/dof = 1467/1291.
Note the Si K$\alpha$ line was modeled separately with a single Gaussian.
The fitted temperature is $kT$ = $0.7^{+0.1}_{-0.1}$ keV, with a high Hydrogen
density $n_H = 3.6^{+11.5}_{-3.0} \times 10^{13} cm^{-3}$. Figure 5 shows the
results of the emission lines fitted by this model in the Mg and Si band.
Obviously, the He-like lines of Si XIII are fitted well with this model.
However, the model over-predicts the emission lines in the Mg band, especially
for Mg XII Ly$\alpha$. Besides it fails to account for the Ne X RRC
(Figure 5(a)) and the Si XIV Ly$\alpha$ line (Figure 5(b)). For comparison, Figure 6 shows that
a photoionization model fits better than the collisional model to the emission
lines for both the He-like and H-like lines, especially for the RRC feature.
Therefore, a collisional origin appears to be excluded for
the soft X-ray emission lines. This is consistent with the diagnostics
of the LETG spectrum using the O VII and O VIII lines (Reeves et al.~\cite{r10}).

  \begin{figure}
   \centering
   \includegraphics[width=12cm, angle=0]{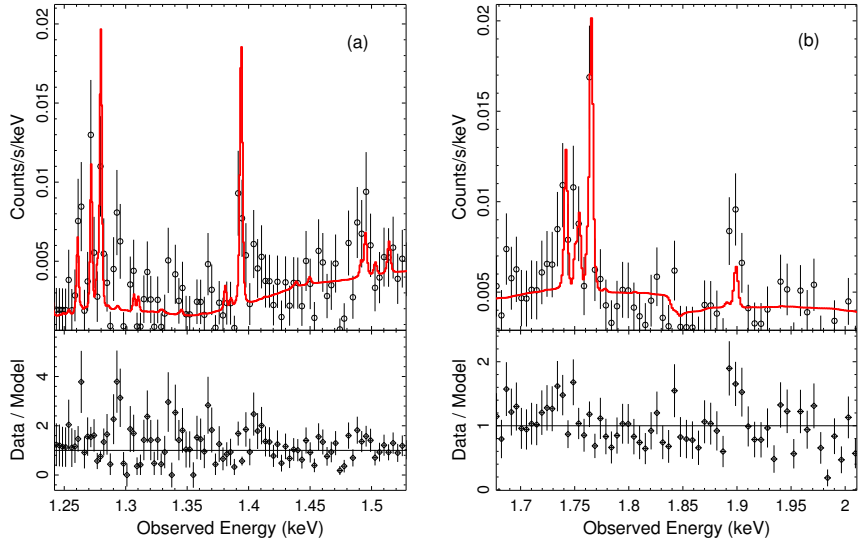}
   \caption{The emission lines fitted with a collisional model. Panel (a) shows the spectrum in the Mg band.
             Panel (b) shows the spectrum in the Si band.}
   \label{Fig5}
   \end{figure}

\subsection{Photoinization Model}

In order to extract the physical conditions in the line emitting gas,
we attempted to model the emission lines using the photoionization
code CLOUDY C13.05 (Ferland et al.~\cite{f13}) in this section. A
plane-parallel geometry is assumed, with the slab depth controlled by
the hydrogen column density parameter ($N_H$). The free parameters
in the model are: the ionization parameter ($\xi$), the density of
the material ($n_H$) and column density ($N_H$) of the gas.

  \begin{figure}
   \centering
   \includegraphics[width=9cm, angle=0]{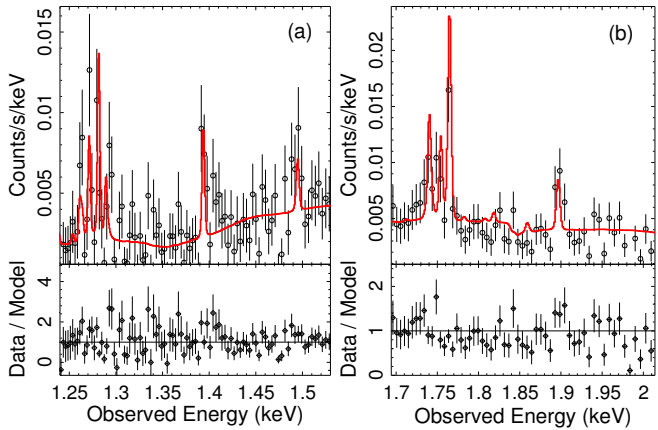}
   \caption{The emission lines fitted with a photoionization model. Panel (a) shows the spectrum in the Mg band.
           Panel (b) shows the spectrum in the Si band.}
   \label{Fig6}
   \end{figure}

As mentioned in the above section, the He-like emission line intensities
critically depend on the spectral energy distribution (SED) of the
ionizing continuum, especially the UV field. However, determining the SED
is difficult since the line emitting gas might see a very different
continuum from what we see (Korista et al.~\cite{k97}). In the present
calculation, we assumed a typical AGN continuum (i.e. the model AGN in
CLOUDY), which is defined by a "big bump" with temperature $T$ = $1.5 \times 10^5$ K
and a low-energy slope of $\alpha_{UV}$ = -0.5, a typical value of X-ray
to UV ratio $\alpha_{OX}$ = -1.4, and an X-ray power-law of spectral index
of $\alpha$ = -1. As discussed in Section 3.2, the density of the material
is high and it was taken in the range of $n_H = 10^8 - 10^{13} cm^{-3}$
in our calculation.
The low values of the G ratio indicate a low column density for the gas
(Porter \& Ferland~\cite{p07}), and we adopted the value in the
range of $N_H = 10^{19} - 10^{23} cm^{-2}$. The ionization parameter is set
in the range of log{$\xi$} (erg cm s$^{-1}$) = 1 - 4. Based on the width of the
Ne X RRC, the temperature was fix to $kT_e$ = 3 eV in the calculation.
In the analysis of LETG spectrum of 3C445, Reeves et al.~(\cite{r10}) found some tentative
evidence for super-solar abundances of Mg and Si (2.6 and 6.5 times of the
solar value respectively), and this is confirmed by the Suzaku observations (Braito et al.~\cite{b11}).
Thus we assumed the same abundances for Mg and Si with
previous researches, while the other abundances were fixed to the solar value.

A grid of emission line spectrum were generated and incorporated into the
partial covering model, and fitted to the overall spectra. The Si K$\alpha$
line was still modeled with a single Gaussian. This model provides a better
fit to the HETG spectra than the collisional model, with the final overall
fit statistic $C$/dof = 1454/1290. When adding a second photoionized
component to the overall spectra, the statistic only improves by
$\Delta{C}$ = 1 for 4 parameters. This suggests a second component
is not formally required. Thus the results are obtained with only
one photoionized component. Figure 6 shows that the model gives a reasonable
fit to the emission lines in the Mg and Si band.

\section{Results and Discussion}
\label{sect:discussion}
The overall fitted spectrum with
the photoionzation model is shown in Figure 7, and the
best-fit parameters are listed in Table 2. Parameters for the continuum are
basically consistent with previous measurements with LETG or Suzaku data
(Reeves et al.~\cite{r10}, Braito et al.~\cite{b11}), except for the photon
index of the primary power-law. The value $\Gamma$ = $1.4^{+0.3}_{-0.3}$ is
lower than the typical value for BLRG objects ($\Gamma$ = $1.74^{+0.3}_{-0.3}$)
(Grandi et al.~\cite{g04}), indicating a hard/flat continuum. This is
probably due to low quality of the data in the soft X-ray band. Given the
uncertainty in the value, the photon index is roughly consistent with the typical value
for BLRGs.

\begin{table}
\begin{center}
\begin{threeparttable}
\caption[]{Summary of the Overall Spectra Model Parameters.}\label{Tab:2}
\begin{tabular}{ccc}
  \hline\noalign{\smallskip}
 Model Name & Parameter & Value \\
 \hline\noalign{\smallskip}
 Power-law           & $\Gamma$                  & $1.4^{+0.3}_{-0.3}$\\
                     & Normalization             & $2.7^{+2.3}_{-1.3} \times 10^{-3}$\\
 Scattered Power-law & Normalization             & $7.2^{+1.0}_{-1.1} \times 10^{-5}$ \\
 Ionizaed reflection &  $\xi$\tnote{a}           & 10 erg cm s$^{-1}$ \\
                     & Normalization             & $1.2^{+0.3}_{-0.3} \times 10^{-5}$\\
 Absorber1           & $N_H$                     & $6.4^{+1.3}_{-1.3} \times 10^{22} cm^{-2}$\\
 Abserber2           & $N_H$                     & $2.4^{+0.7}_{-0.6} \times 10^{23} cm^{-2}$\\
                     & CvrFract\tnote{b}         & $0.76^{+0.06}_{-0.09}$\\
 photoionization model & log$\xi$\tnote{c}       & $3.3^{+0.4}_{-0.3}$ \\
                      & $N_H$                    & $2.5^{+3.8}_{-1.7}\times10^{20}$ cm$^{-2}$\\
                      & $n_H$                    & $5^{+15}_{-4.5}\times10^{10}$ cm$^{-3}$ \\
\noalign{\smallskip}\hline

\end{tabular}
\begin{tablenotes}
\item[a]{The ionization parameter of the reflection component, fixed to the minimum value in the overall fitting.}
\item[b]{Covering factor of the second absorbing component.}
\item[c]{The ionization parameter in units of erg cm s$^{-1}$.}
\end{tablenotes}
\end{threeparttable}
\end{center}
\end{table}

According to the ionization parameter, the photoionized component detected here corresponds
to the second component detected in the LETG spectrum (i.e. the higher-ionization
zone, see Reeves et al.~\cite{r10}). A small column density
$N_{H}$ = $2.5^{+3.8}_{-1.7}\times10^{20}$ cm$^{-2}$ was found for the
line emitting gas, which is consistent with the values derived from several recent
high-resolution soft X-ray spectra of the Seyfert galaxies (e.g. Mrk 573,
Gonzalez-Martin et al.~\cite{gm10}; Mrk 335, Longinotti et al.~\cite{l08} or
NGC 1068, Kraemer et al.~\cite{k15}). A high Hydrogen density was found with
$n_{H}$ = $5^{+15}_{-4.5}\times10^{10}$ cm$^{-3}$, which is consistent with
the analysis on the R ratios of the Mg XI and Si XIII He-like triplets in Section 3.2.

  \begin{figure}
   \centering
   \includegraphics[width=10cm, angle=0]{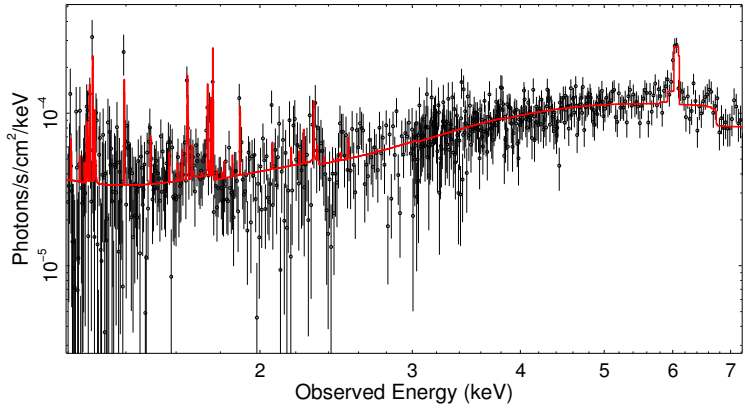}
   \caption{The overall MEG + HEG spectrum fitted by the model described in Section 4.}
   \label{Fig7}
   \end{figure}

With the best-fit parameters, we can estimate the distance of the gas to the
central source. The distance $R$ can be obtained by the definition of the ionization
parameter, i.e. $\xi$ = $L_{ion}/{{n_H}{R^2}}$ (Tarter et al.~\cite{t69}),
where $L_{ion}$ is the ionization luminosity in 1 - 1000 Ryd, and $n_H$ is the
Hydrogen density. Based on the SED used here, we calculated the ratio between the
ionization luminosity and the bolometric luminosity, i.e. luminosity ratio between 1 - 1000 Ryd
range and the whole energy range. Using the bolometric luminosity of 3C445
($L_{bol}$ = $1.3 \times 10^{45}$ erg/s, Grandi et al.~\cite{g06}), the ionization luminosity
was determined $L_{ion}$ = 3.6 $\times$ $10^{44}$ erg/s. Then the radial distance
was derived as $R$ $\sim$ $1.9 \times 10^{15}$ cm (6 $\times$ $10^{-4}$ pc).
Considering the uncertainties in the parameters, the distance was ultimately
determined lying in the range of 6 $\times$ $10^{14}$ - 8 $\times$ $10^{15}$ cm.
Interestingly, the range is well within the optical BLR of 3C445 with
$R$ = 1.2 $\times$ $10^{16}$ cm (Osterbrock et al.~\cite{o76}),
suggesting we might be viewing the inner part of the BLR.
Indeed, the best-fit Hydrogen density is consistent with the typical value of
the BLR in AGN ($10^{8} - 10^{12} cm^{-3}$; Gonzalez-Martin et al.~\cite{gm10},
Longinotti et al.~\cite{l10}).
Besides, as discussed in Section 3, the FWHM ratio of Si XIV and H$_{\beta}$ is
consistent with 1 at the two-parameter 99$\%$ confidence level, which also supports that
the soft X-ray emitting gas locates in the BLR.

As shown in Figure 7, the strong Fe K$\alpha$ line can be well modeled by the reflection
component ($reflion$), confirming the line originating from reflection off the
Compton-thick matter detected in previous observations (Grandi et al.~\cite{g06}, Braito et al.~\cite{b11}).
We estimated the location of this line emitting gas using its measured line width,
i.e. $v_{FWHM}$ $\sim$ $4100^{+2300}_{-1400}$ km/s measured in the HEG spectrum.
Assuming Keplerian motion, the distance of the gas to the central engine
can be estimated as R = GM/$v^2$, where M is the mass of the central black hole in
3C445 (which has an estimated value of 2 $\times$ $10^8$ M$_{\odot}$; Grandi et al.~\cite{g06}).
The velocity dispersion $v$ is related to FWHM velocity as $v^2$ = $\frac{3}{4}$$v_{FWHM}^2$.
Thus we inferred the distance R $\sim$ 2 $\times$ $10^{17}$ cm, comparable with the size of the optical BLR
($R$ = 1.2 $\times$ $10^{16}$ cm; Osterbrock et al.~\cite{o76}).
This indicates that the Fe K$\alpha$ line probably originates from the BLR or outer region
of the accretion disk. Using the same method, we estimated the distance of the Si K$\alpha$ line
emitting gas R $>$ 2 $\times$ $10^{18}$ cm ($>$ 1 pc), which is consistent with the putative
parsec-scale torus.

Our best-fit model seems to confirm the inner geometry of 3C445 as proposed in Reeves et al.(~\cite{r10}).
The central primary X-ray source is heavily absorbed by clumped and neutral/low-ionized matter
(with column density of $\sim$ $10^{23}$ cm$^{-2}$), which may extend to the putative torus.
Conversely, the soft X-ray or optical line emitting clouds are unobscured by the absorber, which
is most probably due to these clouds are lifted above the plane of the accretion disk.
Thus the clouds are photoionized by the central source and generate soft X-ray or optical emission lines.
Meanwhile it scatters the primary X-ray continuum into our line of sight (the scattered power-law in Table 2).
This structure is very similar to that observed in the Seyfert 2 galaxies. One of the most notable example
is the Seyfert 2 galaxy Mkn 3, in which broad optical emission lines are observed in polarized light
(Miller \& Goodrich~\cite{mg90}; Tran~\cite{t95}), suggesting the BLR in this object is scattering the primary
radiation into our line of sight. The main difference is that we are viewing directly the BLR of 3C445,
while the BLR is obscured in Mkn 3.

\section{Summary}
\label{sect:conclusion}
Using the high resolution Chandra HETG spectrum, we have analyzed the soft
X-ray emission for the BLRG object 3C445. Highly ionized emission features
are significantly detected and identified from H- or He-like Ne, Mg, Si and S ions.
A prominent neutral Si K$\alpha$ line was also identified in the spectrum,
which was supposed originating from Compton-thick matter existing in 3C445.
we measured the centroid energy, fluxes, widths, and the equivalent widths of
these features, and modelled the spectrum with a collisional model and a
photoionization model, respectively. The main results are as follows:

 1. Most of the features are detected with the confidence level $>$ $99\%$.
    Most of the measured line centroid energy are consistent with the experimental values,
    suggesting no velocity shift is required for the emitter. The measured line FWHMs are
    consistent with the H$_{\beta}$ FWHM (3000 km/s) at two-parameter 99$\%$ confidence level.
    The high values of the R ratios (R = $0.9^{+1.1}_{-0.5}$ and $1.3^{+2.3}_{-0.7}$ for Mg XI
    and Si XIII respectively) suggest a high density for the line emitting matter.

 2. A relatively weak Ne X RRC feature is detected in the spectrum.
    The narrow RRC feature indicates the electron temperature of the line emitter is low
    (with $kT_e$ $<$ 4 eV). This supports a photoionization origin for the highly-ionzed emission lines.

 3. The G ratios of He-like lines indicate the emission lines originate from collisionally
    ionized plasmas. However, modelling the spectrum with a collisional ionization model fails to account for
    the emission lines present in the spectrum. This excludes a collisional origin for the emission lines.

 4. A photoionization model succeeded in modelling the spectrum, with the best-fit parameters
    log$\xi$ = $3.3^{+0.4}_{-0.3}$ erg cm s$^{-1}$, $n_{H}$ = $5^{+15}_{-4.5}\times10^{10}$ cm$^{-3}$
    and $N_{H}$ = $2.5^{+3.8}_{-1.7}\times10^{20}$ cm$^{-2}$. The high density of the emitter, the line
    FWHMs and the inferred radial distance all suggest the emitter locates in the BLR in 3C445.

\begin{acknowledgements}
We gratefully acknowledge the anonymous referee for the helpful suggestions that improve the performance
of this manuscript.
This work is based on observations made by the Chandra X-ray Observatory, and has made use of the CIAO
software provided by the Chandra X-ray Center.

\end{acknowledgements}

\label{lastpage}


\begin{thebibliography}{99}

\bibitem[2007]{a07}Armentrout, B. K., Kraemer, S. B., Turner, T. J. 2007, ApJ, 665, 237
\bibitem[2006]{b06}Bianchi, S., Guainazzi, M., Chiaberge, M. 2006, A\&A, 448, 499
\bibitem[2011]{b11}Braito, V., Reeves, J. N., Sambruna, R. M., Gofford, J. 2011, MNRAS, 414, 2739
\bibitem[1979]{c79}Cash, W. 1979, ApJ, 228, 939
\bibitem[2001]{d01}Dere, K. P., Landi, E., Young, P. R., Del Zanna, G. 2001, ApJS, 134, 331
\bibitem[1990]{dl90}Dickey, J. M., Lockman, F. J. 1990, ARA\&A, 28, 215
\bibitem[2004]{eh04}Eracleous, M., Halpern, J. P. 2004, ApJS, 150, 181
\bibitem[2013]{f13}Ferland, G. J., Porter, R. L., van Hoof, P. A. M., et al. 2013, RMxAA, 49, 137
\bibitem[2001]{f01}Freeman, P., Doe, S.,Siemiginowska, A. 2001, \procspie, 4477, 76
\bibitem[2006]{f06} Fruscione, A., et al. 2006, \procspie, 6270
\bibitem[1969]{g69} Gabriel, A. H., Jordan, C. 1969, MNRAS, 145, 241
\bibitem[1973]{g73} Gabriel, A. H., Jordan, C. 1973, ApJ, 186, 327
\bibitem[2010]{gm10} Gonzalez-Martin, O., Acosta-Pulido, J. A., Perez Garcia, A. M.,  Ramos Almeida, C. 2010, ApJ, 723, 1748
\bibitem[2004]{g04} Grandi, P., et al. 2004, IAUS, 222, 101
\bibitem[2006]{g06} Grandi, P., Malaguti, G., Fiocchi, M. 2006, ApJ, 642, 113
\bibitem[2007]{g07} Grandi, P., Guainazzi, M., Cappi, M., Ponti, G. 2007, MNRAS, 381, 21
\bibitem[2003]{g03} Gu, M. F. 2003, ApJ, 582, 1241
\bibitem[2007]{gb07} Guainazzi, M. \& Bianchi, S. 2007, MNRAS, 374, 1290
\bibitem[2001]{k01} Kaspi, S., Brandt, W. N., Netzer, H., et al. 2001, ApJ, 554, 216
\bibitem[2002]{k02} Kinkhabwala, A., Sako, M., Behar, E., et al. 2002, ApJ, 575, 732
\bibitem[1986]{k86} Kronberg, P. P., Wielebinski, R., Graham, D. A. 1986, A\&A, 169, 63
\bibitem[1997]{k97}Korista, K., Ferland, G., Baldwin, J. 1997, ApJ, 487, 555
\bibitem[2008]{k08} Kraemer, S. B., Schmitt, H. R., Crenshaw, D. M. 2008, ApJ, 679, 1128
\bibitem[2015]{k15}Kraemer, S. B., Sharma, N., Turner, T. J., George, Ian M., Crenshaw, D. M. 2015, ApJ, 798, 53
\bibitem[1995]{lo95}Liedahl, D. A., Osterheld, A. L., Goldstein, W. H. 1995, ApJL, 438, 115
\bibitem[1996]{lp96} Liedahl, D. A., Paerels, F. 1996, ApJ, 468, 33
\bibitem[1999]{l99} Liedahl, D. A. 1999, The X-Ray Spectral Properties of Photoionized Plasma and Transient Plasmas, LNP, 520, 189
\bibitem[2008]{l08}Longinotti, A. L., Nucita, A., Santos-Lleo, M., Guainazzi, M. 2008, A\&A, 484, 311
\bibitem[2010]{l10}Longinotti, A. L., et al. 2010, A\&A, 510, A92
\bibitem[1986]{m86}Mewe, R., Lemen, J.R., van den Oord, G. H. J. 1986, A\&AS, 65, 511
\bibitem[1990]{mg90}Miller, J. S., Goodrich, R. W. 1990, ApJ, 355, 456
\bibitem[2010]{n10} Nucita, A. A., Guainazzi, M., Longinotti, A. L., Santos-Lleo, M., Maruccia, Y., Bianchi, S. 2010, A\&A, 515, 47
\bibitem[1976]{o76}Osterbrock, D. E., Koski, A. T., Phillips, M. M. 1976, ApJ, 206, 898
\bibitem[2008]{p08}Piconcelli, E., Bianchi, S., Miniutti, G., et al. 2008, A\&A, 480, 671
\bibitem[2000]{pd00}Porquet, D., Dubau, J. 2000, A\&AS, 143, 495
\bibitem[2001]{p01}Porquet, D., Mewe, R., Dubau, J., Raassen, A. J. J., Kaastra, J. S. 2001, A\&A 376, 1113
\bibitem[2007]{p07}Porter, R. L., Ferland, G. J. 2007, ApJ, 664, 586
\bibitem[2010]{r10} Reeves, J. N., Gofford, J., Braito, V., Sambruna, R. M. 2010, \apj, 725, 803
\bibitem[2007]{s07} Sambruna, R. M., Reeves, J. N., Braito, V. 2007, \apj, 665, 1030
\bibitem[2010]{s10} Shu, X. W., Yaqoob, T., Wang, J. X. 2010, ApJS, 187, 581
\bibitem[2011]{s11} Shu, X. W., Yaqoob, T., Wang, J. X. 2011, ApJ, 783, 147
\bibitem[1969]{t69} Tarter, C. B., Tucker, W. H., Salpeter, E. E. 1969, ApJ, 156, 943
\bibitem[2009]{t09} Torresi, E., Grandi, P., Guainazzi, M., Palumbo, G. G. C., Ponti, G., Bianchi, S. 2009, A\&A, 498, 61
\bibitem[1995]{t95} Tran, H. D. 1995, ApJ, 440, 565
\bibitem[2001]{y01} Young, A. J., Wilson, A. S., Shopbell, P. L. 2001, ApJ, 556, 6


\end{thebibliography}
\end{document}